\documentclass[conference]{IEEEtran}
\usepackage{xcolor,soul,framed} 
\usepackage{enumerate} 
\usepackage{braket}
\usepackage{hyperref}
\usepackage{xurl}
\colorlet{shadecolor}{yellow}
\usepackage[pdftex]{graphicx}
\graphicspath{{../pdf/}{../jpeg/}}
\DeclareGraphicsExtensions{.pdf,.jpeg,.png}
\usepackage{bbold}
\usepackage{algorithm}
\usepackage{algpseudocode}
\usepackage[T1]{fontenc}

\usepackage[cmex10]{amsmath}
\usepackage{array}
\usepackage{mdwmath}
\usepackage{mdwtab}
\usepackage{eqparbox}
\usepackage{amsfonts}
\usepackage{booktabs} 
\usepackage{multirow} 
\usepackage{amsmath}
\usepackage{comment}
\usepackage{amsthm}

\newtheorem{remark}{Remark}

\begin{document}
\bstctlcite{IEEEexample:BSTcontrol}
    \title{To Purify or Not to Purify: Entanglement Purification under Input Fidelity Asymmetry in Quantum Networks}
  \author{Anoosha Fayyaz,  Prashant Krishnamurthy, Kaushik Seshadreesan, Amy Babay, and David Tipper\\ 
  \textit{Department of Informatics and Networked Systems} \\
\textit{University of Pittsburgh, Pittsburgh, USA} \\ \{anf224, prashk, kausesh, babay, dtipper\}@pitt.edu
  } 

\maketitle
\begin{abstract}
Entanglement purification with two entangled resource pairs is widely employed in the literature on quantum repeater networks to counteract fidelity degradation introduced by noisy quantum memories and entanglement swapping across multiple hops. Standard purification protocols assume both resource pairs carry identical fidelity. In practice, entanglement generation is stochastic, the two resource pairs are heralded at different times, and so the first pair decoheres in memory while the second is being generated. Thus a fidelity asymmetry is a structural feature of any network operating under realistic memory conditions leading to the question --- when is it beneficial to perform purification? We derive a closed-form \textit{fidelity asymmetry tolerance} $\delta$(F) that governs whether a purification attempt is beneficial. We determine a universal upper bound $\delta_{max} \approx 0.076$ beyond which purification is always counterproductive. Our simulations show that with exponential memory decoherence, purification yields benefits in only approximately 14\% of purification attempts on two resource pairs in a two-hop repeater chain. We define three network objectives --- fidelity only, time only, and a combination of time and fidelity --- to deliver end-to-end entanglement. Under the memory models and network assumptions studied, we show that when the application fidelity requirement is achievable through swapping alone, no-purification is the superior policy, with its advantage increasing with the number of hops. When the fidelity requirement cannot be met with swapping alone and purification is necessary, to be effective, it must be conditioned on $\delta$(F) between resource pairs. We introduce DeltaPurify, a policy that conditions purification decisions on local fidelity information, and show it reduces time-to-serve relative to both naive purification and no-purification across several fidelity thresholds and hops of a repeater chain.
\end{abstract}

\section{Introduction}
\label{sec:introduction}

\IEEEPARstart{Q}{uantum} networking has advanced rapidly in recent years, with experimental demonstrations moving beyond the laboratory. Testbeds operating over deployed metropolitan fiber now exist in several cities. QuTech demonstrated qubit teleportation across a three-node network \cite{hermans2022qubit}, Harvard and AWS entangled quantum memory nodes over 35~km of fiber beneath the streets of Boston \cite{knaut2024entanglement}, and EPB launched the first commercially available quantum network in Chattanooga \cite{epb2022quantum}. Most recently, Qunnect and Cisco demonstrated entanglement swapping over 17.6~km of deployed fiber in New York City, achieving record generation rates on commercial infrastructure \cite{craddock2026high}. These demonstrations mark a clear shift in quantum networking evolving from a laboratory prototype to one that needs engineering. Consequently, the design decisions that governed controlled experiments must now be re-examined under real-world conditions.

A quantum network distributes entanglement between distant nodes by dividing long links into shorter elementary segments connected by quantum repeaters. Each repeater stores qubits locally, generates entanglement with its neighbors, and stitches together adjacent links through entanglement swapping to progressively extend the end-to-end range. At each stage, two categories of imperfection degrade the quality of the distributed entangled state, measured by its fidelity $F$ to the target state. The first is \textit{structural noise}, in which gate imperfections, detector inefficiencies, and multi-photon emission introduce a baseline fidelity loss each time an entanglement generation or swap operation is performed. The second is \textit{temporal noise}, which arises due to decoherence in quantum memories. Because entanglement generation is probabilistic with a geometric distribution, with success heralded, different links succeed at different times. Qubits that succeed early must wait in memory for neighboring links to complete, and during this interval, they decohere. The longer the wait, the lower the fidelity. Both imperfections are irreversible and compound across hops, so managing them is the central engineering challenge of any quantum repeater network.

\begin{figure}[t]
    \centering
    \includegraphics[width=1\linewidth]{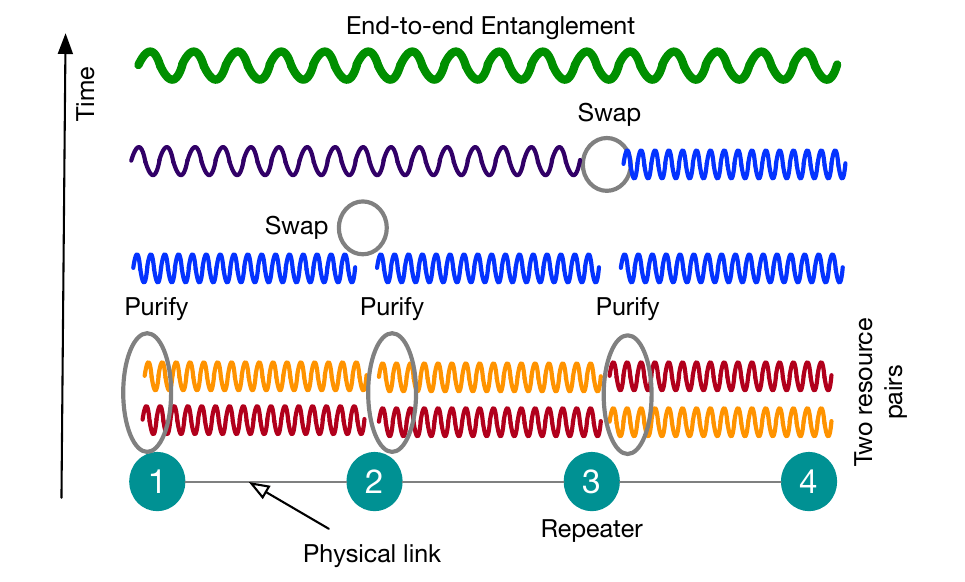}
    \includegraphics[width=1\linewidth]{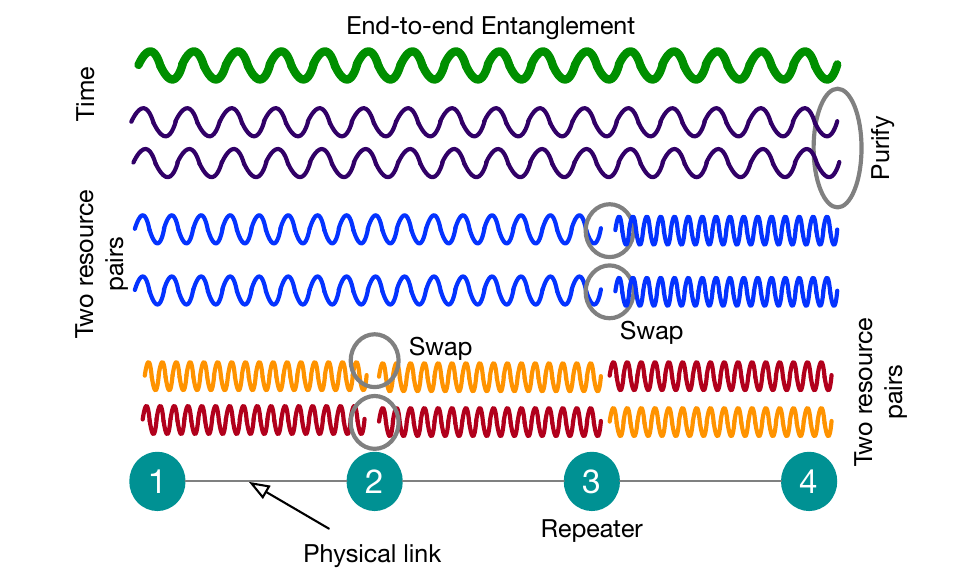}
    \caption{Schematic of a quantum repeater chain with three hops and two purification policies. (Top) In the Purify-Swap (PS) policy, purification is performed at the elementary link level before swapping. (Bottom) In the Swap-Purify (SP) policy, purification is applied to end-to-end pairs after all swaps are complete (bottom). The decision of whether to purify at all is not shown as a separate policy in this schematic, but is evaluated in the results section.
    }
    \label{fig:schematic}
\end{figure}

Entanglement purification is the standard active countermeasure to improve fidelity. A purification protocol consumes $k \geq 2$ low-fidelity resource pairs (entanglements) and, upon success, produces a single entanglement pair of higher fidelity. The seminal BBPSSW \cite{bennett1996purification} and DEJMPS \cite{deutsch1996quantum} protocols established the theoretical foundations, and subsequent work has refined them for practical network settings. The principal scheduling question is \textit{where} in the repeater chain should purification be applied relative to swapping. The Purify-Swap (PS) strategy applies purification at the elementary (physical) link level before swapping, while the Swap-Purify (SP) strategy first establishes multiple end-to-end entanglement between a pair of quantum nodes and then purifies the resulting long-range pair (see Figure~\ref{fig:schematic}). Hybrid strategies that apply purification at intermediate ``virtual" link lengths have also been studied \cite{dur1999quantum, hu2024high}.

\begin{figure}[h!]
    \centering
    \includegraphics[width=0.8\linewidth]{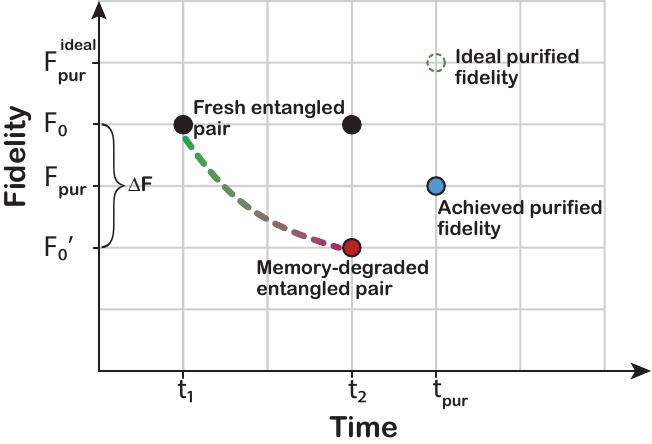}
    \caption{Fidelity decay under temporal decoherence and its consequence for purification. A pair heralded at $t_1$ enters memory at fidelity $F_0$ and decays to $F_0'$ while awaiting a second resource pair heralded at $t_2$. At $t_{\text{pur}}$, the ideal purified fidelity $F_{\text{pur}}^{\text{ideal}}$ assumes symmetric inputs and exceeds $F_0$, whereas the actual realized fidelity $F_{\text{pur}}$ falls below $F_0$ due to the asymmetry induced by storage.}
    \label{fig:EMM-fid-decay}
\end{figure}

Most research (see Section~\ref{sec:bg}) assumes both resource pairs have the same fidelity either (a) through adoption of a \textit{Constant Memory Model} (CMM) where the fidelity of a stored qubit does not decohere for the time period under consideration or (b) by treating each purification event in isolation from the stochastic generation process that produced them. This assumption is realistically unjustified, and its consequences for purification policy are the central subject of this paper. Under a CMM, qubits stored in quantum memories retain their fidelity indefinitely until used or discarded. Two resource pairs therefore have identical fidelity regardless of when their respective entanglements were heralded. Thus, purification is unconditionally beneficial, since the output fidelity always exceeds the input and the option of a single end-to-end entanglement delivery is rendered inferior in terms of fidelity.  
This assumption, however, is optimistic. Real quantum memories decohere continuously, and the relevant figure of merit is not the fidelity at generation but the fidelity at the moment of use. 
As Fig.~\ref{fig:EMM-fid-decay} shows, a pair generated at $t_1$ and stored until $t_2$ decays from $F_0$ to $F_0'$. The ideal purified fidelity $F_{\text{pur}}^{\text{ideal}}$, computed under symmetric inputs at $F_0$, exceeds $F_0$, whereas the actual realized fidelity $F_{\text{pur}}$ falls below $F_0$ due to the asymmetry induced by storage. When this asymmetry is severe enough, the purification operation does not improve the state; it corrupts the higher-fidelity pair by mixing it with the already-degraded one, producing an output of lower quality than the best available input. In that regime, the correct decision is to not purify. This fidelity asymmetry is not an edge case; it is the generic condition in any network operating under a time-dependent decoherence model. Moreover, the time at which a link is heralded and the fidelity it carries are jointly determined by $p_e$, $p_s$, and the memory decoherence model. A purification policy that does not account for these joint parameters cannot reliably determine whether purification will be beneficial at the moment it is invoked.

In this work, we analyze the conditions under which entanglement purification is beneficial in a probabilistic quantum repeater chain. We derive a closed-form asymmetry tolerance $\delta(F)$ governing whether purification of two resource pairs at fidelities $F_1,F_2$ yields a net gain, and a universal upper bound $\delta_{\max}$ on this tolerance (Section~\ref{sec:asymmetry}). We also define $F_{\lim}$, an optimistic upper bound on the fidelity achievable through swapping alone, as a guide for when purification is necessary. Across three memory models, namely Constant (CMM), Linear (LMM), and Exponential (EMM), we show that asynchronous heralding drives input asymmetry beyond $\delta(F)$ in the majority of trials, yielding a net gain in only 14--16\% of runs (LMM/EMM) versus 100\% under CMM. We evaluate the resulting policy implications by comparing No-Purification (No-Pur), Swap-Purify (SP), and Purify-Swap (PS) across three network objectives: minimum-latency delivery above a fidelity threshold, maximum-fidelity delivery within a time budget, and joint fidelity-and-time constraints.

Building on this criterion, we introduce DeltaPurify, a policy that checks $F_{\text{th}}$ against $F_{\lim}$ to determine whether purification is necessary, then applies a feasibility check and a $\delta(F)$ check before committing to a purification attempt, restarting generation whenever a check fails. Only the initial $F_{\lim}$ check requires path-level information; all subsequent decisions use only local, per-pair state (Section~\ref{subsec:policy-comparison}).

Together, these results show that the conventional assumption of
unconditionally beneficial purification does not hold under realistic
memory conditions: purification works as theory predicts, but its
conditions for benefit are routinely violated in practice. Under the
linear-chain topologies studied, No-Pur is preferable across all three
objectives when $F_{\text{th}} \leq F_{\lim}$, with its advantage widening with hop count, while purification is necessary but must be conditioned on $\delta(F)$ when $F_{\text{th}} > F_{\lim}$.

The remainder of this paper is organized as follows. Section~\ref{sec:background} provides the technical background on entanglement generation, swapping, and purification, and defines the memory models used throughout. Section~\ref{sec:asymmetry} develops the analytical theory of input fidelity asymmetry and its implications for network control logic. Section~\ref{sec:results} presents the simulation results comparing the three scheduling policies across the three network objectives, and Section~V concludes.

\section{Background}
\label{sec:background}

\subsection{Related work}
\label{sec:bg}

The trade-offs between purification policies have been studied extensively
since D\"ur et al.~\cite{dur1999quantum} established purify-then-swap as the
canonical first-generation repeater architecture. Subsequent work spans
several network objectives: Victora et al.~\cite{victora2023entanglement}
jointly optimize path and protocol choice for pair-production rate and
fidelity; Wang et al.~\cite{wang2023efficient} derive an optimal
swap/purify ordering maximizing rate under fidelity constraints; Jia and
Chen~\cite{jia2024entanglement} give closed-form conditions for PS
optimality and extend to multi-flow routing; Xiao et
al.~\cite{xiao2024purification} cast purification scheduling as
joint-routing throughput maximization; and Koutsopoulos et
al.~\cite{koutsopoulos2024optimal} give polynomial-time swap-purify
coordination algorithms for two-link networks, observing that omitting
purification can help in some regimes. Haldar et
al.~\cite{haldar2025reducing} identify beneficial-distillation regimes but
omit no-purification as a baseline, while Kamin et
al.~\cite{kamin2023exact} and Rozpedek et
al.~\cite{rozpkedek2018parameter} study no-purification hardware regimes as a
fixed architectural choice rather than a policy alternative. None of these
works evaluates a single end-to-end pair as a standalone policy, and whether
to purify at all is never the central question.

Zang et al.~\cite{zang2025entanglement} study the CNOT-based recurrence EPP under non-identical input fidelities and memory decoherence, proving guaranteed improvement over practical baselines, including the average input fidelity, and deriving a parameter-independent optimal invocation time. Their guaranteed improvement is defined relative to the average of the input fidelities rather than the better input, so their result does not address whether purification improves over the best available resource pair, which is the question central to this paper. Their analysis considers purification in isolation and does not extend to policy comparison in a repeater chain setting. 

\subsection{Quantum Definitions}

\noindent\textbf{Entanglement Generation --- } Entanglement generation is the fundamental process of establishing a link-level Bell pair between two adjacent nodes. Regardless of the physical implementation, the probability of successfully generating an entangled pair in a single clock cycle, denoted $p_e$, is constrained by the efficiency of the underlying hardware. For a mid-point heralding scheme such as Barrett-Kok \cite{barrett2005efficient}, this probability is given by
\begin{equation}
\label{eq:p_e_definition}
p_e = \frac{1}{2} \eta_d^2 \cdot \eta_c^2 \cdot e^{-L/L_{att}},
\end{equation}
where $\eta_d$ is the photon detector efficiency, $\eta_c$ is the fiber-to-memory coupling efficiency, $L$ is the distance between nodes, and $L_{att}$ is the attenuation length of the fiber (typically $\sim 22$~km for silica fibers at 1550~nm). We assume symmetric link lengths and uniform hardware specifications, giving a constant $p_e$ across all links. Due to gate infidelities, dark counts, and multi-photon emissions, the generated pair has an imperfect fidelity $F_0 < 1$. We model generated states as Werner states, which are defined by a single parameter $w \in [0,1]$ related to fidelity by $F = (3w+1)/4$. In this work we work directly in terms of $F$; the initial state is thus fully characterized by the fidelity parameter $F_0$, which serves as the input fidelity at $t = 0$ and subsequently undergoes temporal decay according to the memory model defined in Section~\ref{subsec:system-definitions}. For a pair to be useful for swapping or purification, it must satisfy $F_0 > 0.5$; below this bound the state is separable and cannot support quantum communication.

\noindent
\textbf{Entanglement Swap --- } Entanglement swapping is the process by which two independent link-level entangled pairs are combined to form a longer-range entangled pair. Suppose nodes $A$ and $B$ share an entangled pair $\rho_{AB}$ and nodes $B$ and $C$ share an entangled pair $\rho_{BC}$. We define the link fidelities as $F_{AB} = \bra{\phi^+}\rho_{AB}\ket{\phi^+}$ and $F_{BC} = \bra{\phi^+}\rho_{BC}\ket{\phi^+}$, where all fidelities are defined with respect to the $\ket{\phi^+}$ Bell state. Performing a Bell-state measurement (BSM) on the two qubits at node $B$ creates an entangled pair $\rho_{AC}$ between the distant nodes $A$ and $C$. The fidelity of the swapped state $F_{swap} = \bra{\phi^+}\rho_{AC}\ket{\phi^+}$ is given by
\begin{equation}
    \label{eq:swap}
    F_{swap} = \frac{1}{4} + \frac{3}{4} 
    \left(\frac{4F_{AB}-1}{3}\right)\left(\frac{4F_{BC}-1}{3}
    \right).
\end{equation}

A fundamental property of swapping two non-maximally entangled states is that $F_{swap} < \min(F_{AB}, F_{BC})$, meaning the output fidelity is strictly lower than the worst input. To ensure the resulting pair remains entangled ($F_{swap} \geq 0.5$), the input fidelities must satisfy $(4F_{AB}-1)(4F_{BC}-1) \geq 3$, which sets a lower bound on $F_{BC}$ for a given $F_{AB}$ as
\begin{equation}
    \label{eq:fidelity}
    F_{BC} \ge \frac{1}{4}\Big(\frac{3}{4F_{AB}-1} + 1\Big).
\end{equation}

\noindent
\textbf{Entanglement Purification ---  } Purification is a probabilistic process in which a failed attempt discards all participating pairs, and multiple rounds are typically required to reach a target fidelity, making purification resource-intensive in practice. Entanglement purification is a probabilistic protocol $\mathcal{P}: \rho_i^{\otimes k} \to \rho'$ that maps $k$ low-fidelity resource pairs into a single pair of higher fidelity $F'$. Let nodes $A$ and $B$ share $k \ge 2$ entangled pairs $\{\rho_{AB}^{(i)}\}_{i=1}^k$ with fidelities $F_i = \bra{\phi^+}\rho_{AB}^{(i)}\ket{\phi^+}$. Through local operations and classical communication (LOCC), a single output state $\rho_{AB}'$ is produced with distilled fidelity $F' > \max(F_i)$.

Purification protocols are broadly classified by their resource consumption. Nested protocols such as BBPSSW \cite{bennett1996purification} and DEJMPS \cite{deutsch1996quantum} process pairs in discrete rounds, requiring $2^n$ resource pairs for $n$ rounds. We focus on the BBPSSW protocol as the representative purification operation with $n=1$ pair. For two Werner states with fidelities $F_1$ and $F_2$, the output fidelity and success probability are
\begin{equation}
\label{eq:purified-fidelity}
F_{pur}(F_1, F_2) = \frac{F_1 F_2 + \frac{1}{9}(1-F_1)(1-F_2)}{p_{pur}},
\end{equation}

\begin{equation}
\label{eq:purification-success-prob}
p_{pur}(F_1, F_2) = \frac{8}{9}F_1 F_2 - \frac{2}{9}(F_1+F_2) + \frac{5}{9}.
\end{equation}

Entanglement pumping \cite{dur1999quantum} is more resource-efficient, requiring only $n+1$ pairs for $n$ rounds by iteratively improving a single base pair with a sequence of resource pairs. However, pumping converges to a fixed-point fidelity determined by the quality of the resource pairs and cannot exceed this ceiling, whereas nested purification can reach higher fidelities by using previously purified pairs as inputs to subsequent rounds.

\subsection{System Definitions}
\label{subsec:system-definitions}
\noindent
\textbf{Memory Models --- }
\label{subsec:memory-models}
We model fidelity decay in quantum memories using three models. In all cases, stored states are assumed to be Werner states.
\begin{enumerate}[(i)]
    \item \textit{Exponential Memory Model (EMM):} A depolarizing channel acts on each qubit of the stored pair. The Werner parameter $w$ decays exponentially as $w(t) = w_0\, e^{-\Delta t / T_{\mathrm{coh}}}$, where $w_0$ is the initial Werner parameter. Since fidelity and the Werner parameter are related by $F = (3w+1)/4$, this yields a time-dependent fidelity
    \begin{equation}
        \label{eq:depol-memory-fidelity}
        F(t) = \frac{1}{4} + \frac{3}{4} \left(F_0 - \frac{1}{4}\right)e^{-\Delta t/T_{\mathrm{coh}}},
    \end{equation}
    where $T_{\mathrm{coh}}$ is the coherence time constant. Fidelity decays exponentially toward the completely mixed value $F = 1/4$, corresponding to $w = 0$.

    \item \textit{Linear Memory Model (LMM):} Fidelity decreases linearly from $F_0$ toward $0.5$ over the coherence time $T_{\mathrm{coh}}$. In discrete timesteps, this is
    \begin{equation}
    \label{eq:linear-memory-discrete}
        F(t) = F(t-1) - \left(\frac{F_0 - 0.5}{T_{\mathrm{coh}}}\right),
    \end{equation}
    with $F(t)$ clamped to $0.5$ if the decay would reduce it below this value before $T_{\text{coh}}$ timesteps have elapsed.

    \item \textit{Constant Memory Model (CMM):} The stored state maintains fidelity $F_0$ throughout its lifetime and is discarded abruptly at time $\tau$ without gradual decay.
\end{enumerate}

\noindent
\textbf{Swap-purify --- } In the Swap-Purify (SP) architecture, the network generates two end-to-end entanglements before applying purification. For a path of $n$ hops, all $n-1$ intermediate entanglement swaps $\mathcal{S}$ must be heralded to create two independent end-to-end pairs $\{\rho_{e2e}^{(1)}, \rho_{e2e}^{(2)}\}$. The purification operator $\mathcal{P}$ is then applied to these end-to-end states at the terminal nodes. 

\noindent
\textbf{Purify-Swap --- }In the Purify-Swap (PS) architecture, entanglement purification is performed at the elementary link level prior to any 
entanglement swapping. For a node to execute a swap $\mathcal{S}$ between adjacent links $L_i$ and $L_{i+1}$, each link must have first successfully completed the purification map $\mathcal{P}$. Purification is attempted as soon as two entangled pairs on the same link are simultaneously available. Swapping follows a swap-ASAP policy in which a swap is performed as soon as any two adjacent purified links are available in an $n$-hop chain. 
This work restricts its scope to elementary link-level purification and does not consider nested $l$-hop purification where $1 < l < n$.\\

\begin{figure}
    \centering
    \includegraphics[width=0.8\linewidth]{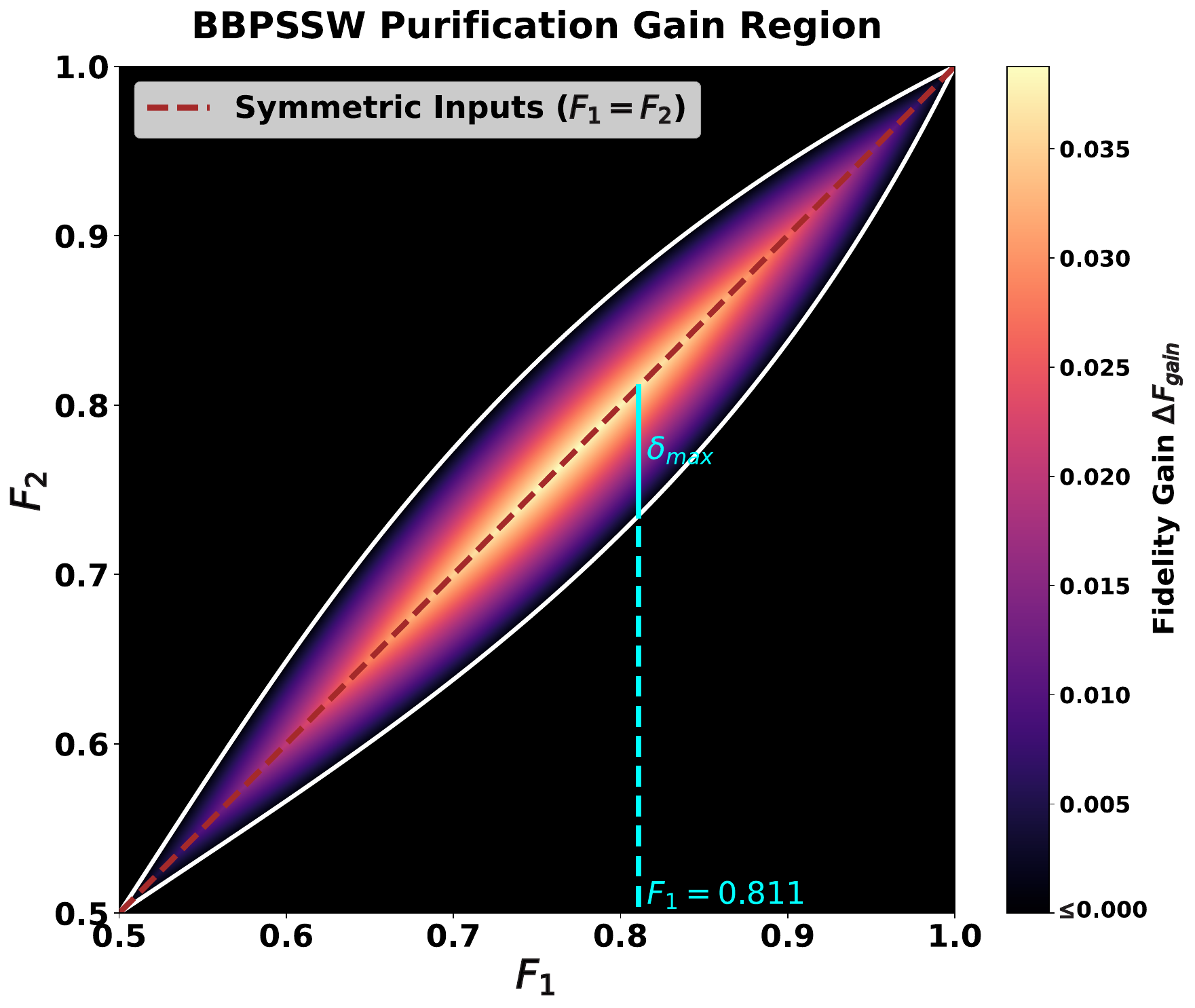}
    \caption{The heatmap illustrates the net fidelity gain, defined as $\Delta F_{\mathrm{gain}} = F_{pur} - \max(F_1, F_2)$, across the range of possible input fidelities $F_1$ and $F_2$. The colored region represents the "purification window" where a successful operation yields a higher fidelity than both inputs. Conversely, the black region indicates the regime where purification is counterproductive, resulting in a state of lower quality than the best available input. The dashed red diagonal denotes the ideal symmetric case ($F_1 = F_2$), where gain is maximized.}
    \label{fig:BBPSSW-pur-gain-region}
\end{figure}

\section{Input Fidelity Asymmetry}
\label{sec:asymmetry}

\subsection{The Symmetry Assumption}

Entanglement purification protocols are designed under the fundamental assumption of input symmetry, namely that both resource pairs are assumed to carry identical fidelity ($F_1 = F_2$) at the moment of purification. Standard frameworks such as BBPSSW and DEJMPS are optimized for this case, and more recent protocols for higher numbers of resource pairs \cite{jansen2022enumerating} likewise assume identical fidelity across all input pairs. In a quantum network, this symmetry is typically maintained by assuming  (i) homogeneous link lengths and hardware specifications, which ensure a uniform initial fidelity $F_0$ across all links and (ii) the Constant Memory Model (CMM) where fidelity remains invariant during storage regardless of wait time in the memory. Together these assumptions ensure both resource pairs arrive at purification with identical fidelity $F_0$.

The CMM assumption is the most consequential to challenge. Under any time-dependent decoherence model, the two resource pairs for a purification operation are heralded at different times. For any two pairs heralded at times $t_1$ and $t_2$, the first must remain in memory for $\Delta t =|t_2 - t_1|$ timesteps while the second is being generated, decaying in fidelity according to Eq.~\ref{eq:depol-memory-fidelity}. Since heralding times are independent and random, the probability of both pairs being heralded in the same timestep is small, and decreases further with the number of hops, making asymmetry increasingly likely for longer-range purification such as SP. A fidelity asymmetry between the freshly heralded pair at $F_0$ and the waiting pair whose fidelity has decayed to $F'$ is therefore present in nearly all purification trials. Once temporal decay is introduced, \textit{input asymmetry is not an edge case but an inevitable feature in a quantum network}.

\subsection{Purification Gain Under Asymmetric Inputs}
\label{subsec:gain-regime}

The fidelity gain $\Delta F_{\text{gain}}$ is defined as the net improvement relative to the best available input,
\begin{equation}
\label{eq:gain_def}
\Delta F_{\text{gain}} = F_{\text{pur}} - \max(F_1, F_2).
\end{equation}
A purification operation is beneficial if and only if $\Delta F_{\text{gain}} > 0$. Figure~\ref{fig:BBPSSW-pur-gain-region} illustrates the gain region for the BBPSSW protocol across the full input fidelity space. The region of positive gain forms a lenticular manifold symmetric about the $F_1 = F_2$ diagonal, confirming that the protocol is optimized for identical inputs. The protocol exhibits a finite tolerance for asymmetry, but the gain diminishes rapidly as $F_1$ and $F_2$ diverge. Operating outside this manifold is counterproductive, as the protocol consumes additional resources to produce a state of lower fidelity than the best available input. The manifold reaches its maximum width, representing the peak asymmetry tolerance, when the higher input fidelity is approximately $F_1 \approx 0.81$, as derived analytically in Section~\ref{subsec:analytical}.

We define the observed input asymmetry as $\Delta F = |F_1 - F_2|$. The following subsection derives a state-dependent tolerance $\delta(F)$ that governs, for any given $F$, the maximum $\Delta F$ the protocol can absorb while still yielding a net gain, reducing the purification decision to a single comparison
\begin{equation}
\label{eq:decision}
\text{purify if and only if } \Delta F < \delta(F).
\end{equation}

\noindent\textbf{Result 1:} (Asymmetry Tolerance). \textit{The BBPSSW protocol yields a net fidelity gain $F_{\mathrm{pur}} > \max(F_1, F_2)$ if and only if $\Delta F < \delta(F)$, where $\delta(F)$ is derived in Section~\ref{subsec:analytical}. The purification decision therefore reduces to evaluating the local criterion of Eq.~\ref{eq:decision}.}

\subsection{Universal Bound on $\delta(F)$}
\label{subsec:analytical}

We now derive the boundaries of the gain region analytically. Let $F_1 \geq F_2$ denote the fidelities of the higher- and lower-fidelity input pairs, respectively. The condition for a net fidelity gain is $F_{\text{pur}} > F_1$. Substituting the BBPSSW output fidelity (Eq.~\ref{eq:purified-fidelity}) and solving for the minimum $F_2$ required to satisfy this condition for a given $F_1$
\begin{equation}
\label{eq:f2min}
F_{2,\min}(F_1) = \frac{2F_1^2 - 6F_1 + 1}{8F_1^2 - 12F_1 + 1}.
\end{equation}
The protocol yields a net fidelity gain if and only if $F_2 > F_{2,\min}(F_1)$. The state-dependent asymmetry tolerance $\delta(F_1)$, representing the maximum observed asymmetry $\Delta F$ the protocol can absorb while still yielding a gain, is therefore
\begin{equation}
\label{eq:delta_local}
\delta(F_1) = F_1 - F_{2,\min}(F_1) = 
\frac{8F_1^3 - 14F_1^2 + 7F_1 - 1}{8F_1^2 - 12F_1 + 1}.
\end{equation}
Note that $\delta(F_1)$ is distinct from the observed asymmetry $\Delta F$. Purification is beneficial if and only if $\Delta F < \delta(F_1)$.

The universal upper bound $\delta_{\max}$ is obtained by maximizing $\delta(F_1)$ over all valid $F_1 \in (0.5, 1)$. Setting $d\delta/dF_1 = 0$ yields a cubic equation in $F_1$ that is solved numerically, giving the critical point $F_1 \approx 0.811$,  $F_2 \approx 0.735$, and
\begin{equation}
\label{eq:delta_max}
\delta_{\max} \approx 0.076.
\end{equation}

\noindent\textbf{Result 2:} (Universal Bound). \textit{For any input pair with $\Delta F > \delta_{\max} \approx 0.076$, purification is counterproductive regardless of absolute fidelity values. For pairs with $\Delta F \leq \delta_{\max}$, viability is governed by the local criterion $\delta(F)$ of Result~1; $\delta_{\max}$ is therefore a necessary
but not sufficient condition for gain.}

The same bound can be expressed from the perspective of the inferior pair. Solving for the maximal permissible $F_1$ given a fixed $F_2$ such that $F_{\text{pur}} > F_1$ is maintained,
\begin{equation}
\label{eq:f1max}
F_{1,\max}(F_2) = \frac{6F_2 - 3 + 
\sqrt{28F_2^2 - 26F_2 + 7}}{2(4F_2-1)}.
\end{equation}

This gives the maximum upward tolerance from the inferior pair: given $F_2$, the superior pair must satisfy $F_1<F_{1, \max}(F_2)$
Setting $d\delta/dF_2 = 0$ recovers the same critical point at $F_2 \approx 0.735$, $F_1 \approx 0.811$, and $\delta_{\max} \approx 0.076$, confirming that the bound is independent of which input serves as the reference.

Importantly, $\delta$ is not symmetric: for a given fidelity value $F$ the maximum upward tolerance when
$F$ is the inferior pair is not equal to the downward tolerance when $F$ is the superior pair. This reflects the asymmetric structure of the BBPSSW gain manifold. Both expressions evaluate the same underlying function $\delta(\cdot)$ but in opposite directions, and together they define the unified purification criterion,
\begin{equation}
\label{eq:delta_unified}
\delta(F) = \begin{cases} 
F_1 - F_{2,\min}(F_1) & \text{if } F = F_1 \\ 
F_{1,\max}(F_2) - F_2 & \text{if } F = F_2 
\end{cases}
\end{equation}

The $\delta(F)$ criterion requires only local fidelity information and translates directly into Eq.~\ref{eq:decision}. The operational consequences are examined in Section~\ref{sec:results}; the direct application in a control policy is presented in Section~\ref{sec:deltapurify}.

\begin{remark}[Protocol independence on Werner inputs]
Although $\delta(F)$ and $\delta_{\max}$ are derived from BBPSSW's output
fidelity (Eq.~\ref{eq:purified-fidelity}), the same expressions govern
DEJMPS~\cite{deutsch1996quantum} whenever both resource pairs are Werner states.
DEJMPS's pre-processing step is a local rotation designed to symmetrize a
general Bell-diagonal state before the bilateral CNOT, but a Werner state
is already invariant under any such rotation (it is invariant under
$U\otimes U^{*}$ for every unitary $U$, not merely on average), so the
rotation acts as the identity and the two protocols reduce to the same
circuit. Consequently, Results~1 and~2 --- including
$\delta_{\max}\approx0.076$ --- hold for DEJMPS without modification on this
input class, and more generally for any two-copy protocol built from a
local rotation followed by bilateral-CNOT measurement and
postselection~\cite{jansen2022enumerating}.
\end{remark}

\begin{figure}[t]
    \centering
    \includegraphics[width=0.8\linewidth]{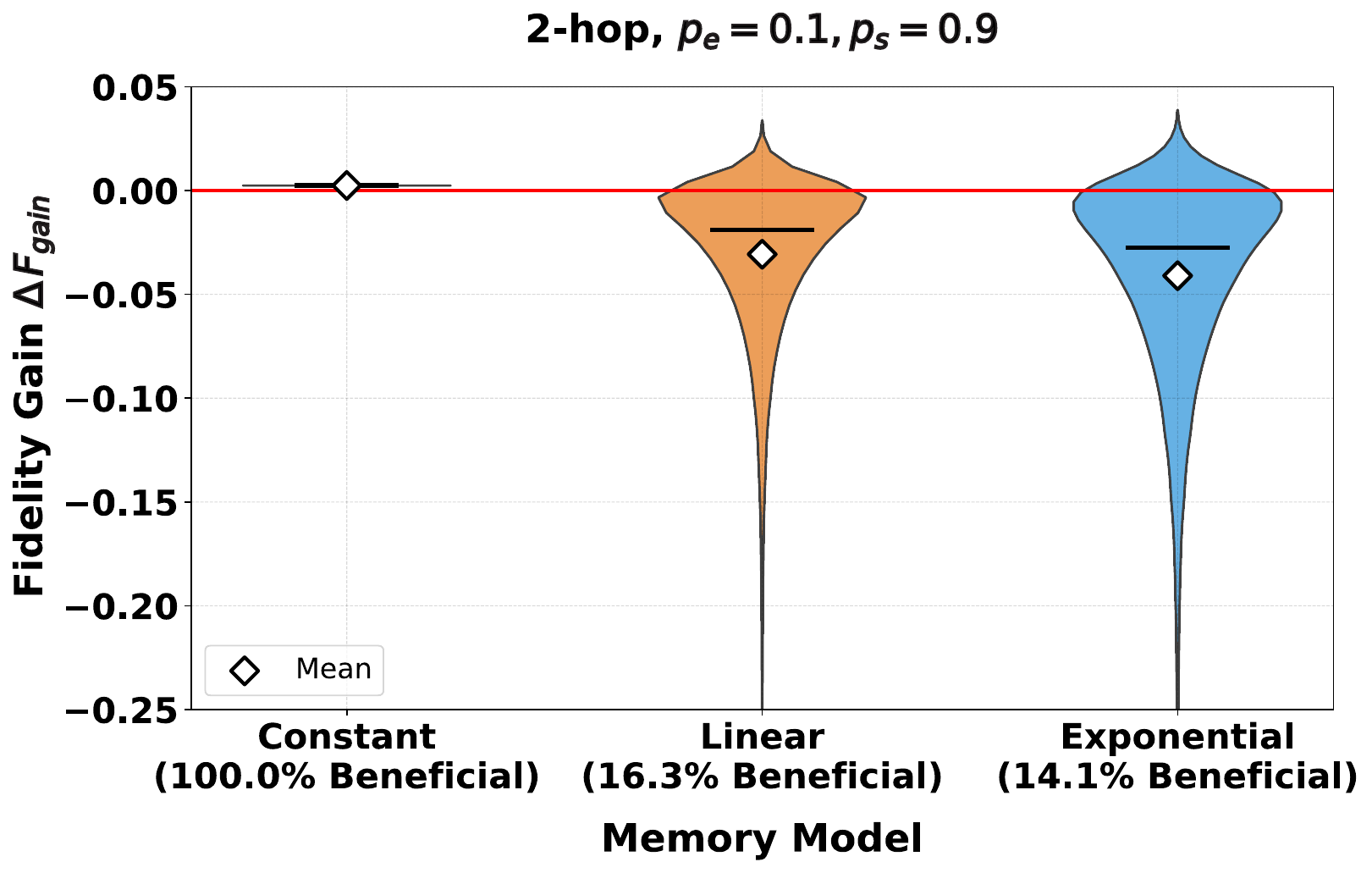}
    \caption{Distribution of fidelity gain $\Delta F_{\text{gain}}$ under three memory models. Under the CMM, purification is unconditionally beneficial. Under the LMM and EMM, the bulk of the distribution lies below zero, indicating that purification actively degrades the best available input in the majority of trials.}
    \label{fig:violin_memory}
\end{figure}

\section{Results}
\label{sec:results}
In this section, we empirically verify the analytical predictions of Section~\ref{sec:asymmetry} and examine their operational consequences across three network objectives. Unless otherwise stated, all results are obtained for a two-hop topology with an initial link fidelity $F_0 = 0.99$, averaged over $10^5$ simulation runs, with $p_e = 0.1$ and $p_s = 0.9$. The qualitative conclusions hold across the broader parameter space $p_e \in [0, 0.5]$, $p_s \in [0.5, 1.0]$, and for lower initial fidelities including $F_0 =0.9$.

\begin{figure*}[t]
    \centering
    \begin{minipage}[t]{0.48\linewidth}
        \centering
        \includegraphics[width=0.8\linewidth]{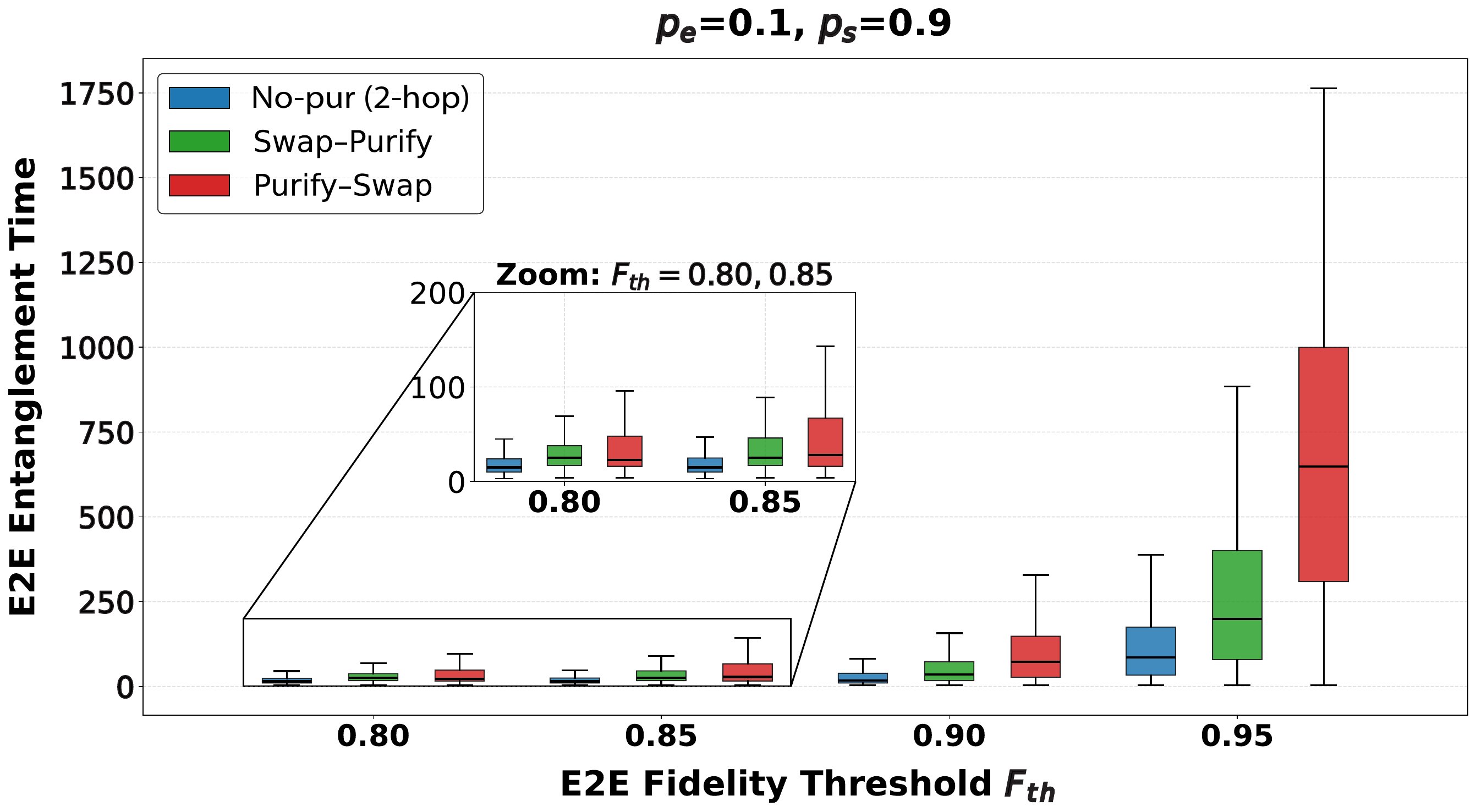}
        \caption{Distribution of end-to-end entanglement delivery times for No-Pur, SP, and PS policies across different fidelity thresholds $F_{th}$. For each threshold value, three stacked boxplots show the delivery time distribution per policy. No-Pur achieves the lowest median and narrowest spread across all threshold values.}
        \label{fig:obj1_latency}
    \end{minipage}
    \hfill
    \begin{minipage}[t]{0.48\linewidth}
        \centering
        \includegraphics[width=0.8\linewidth]{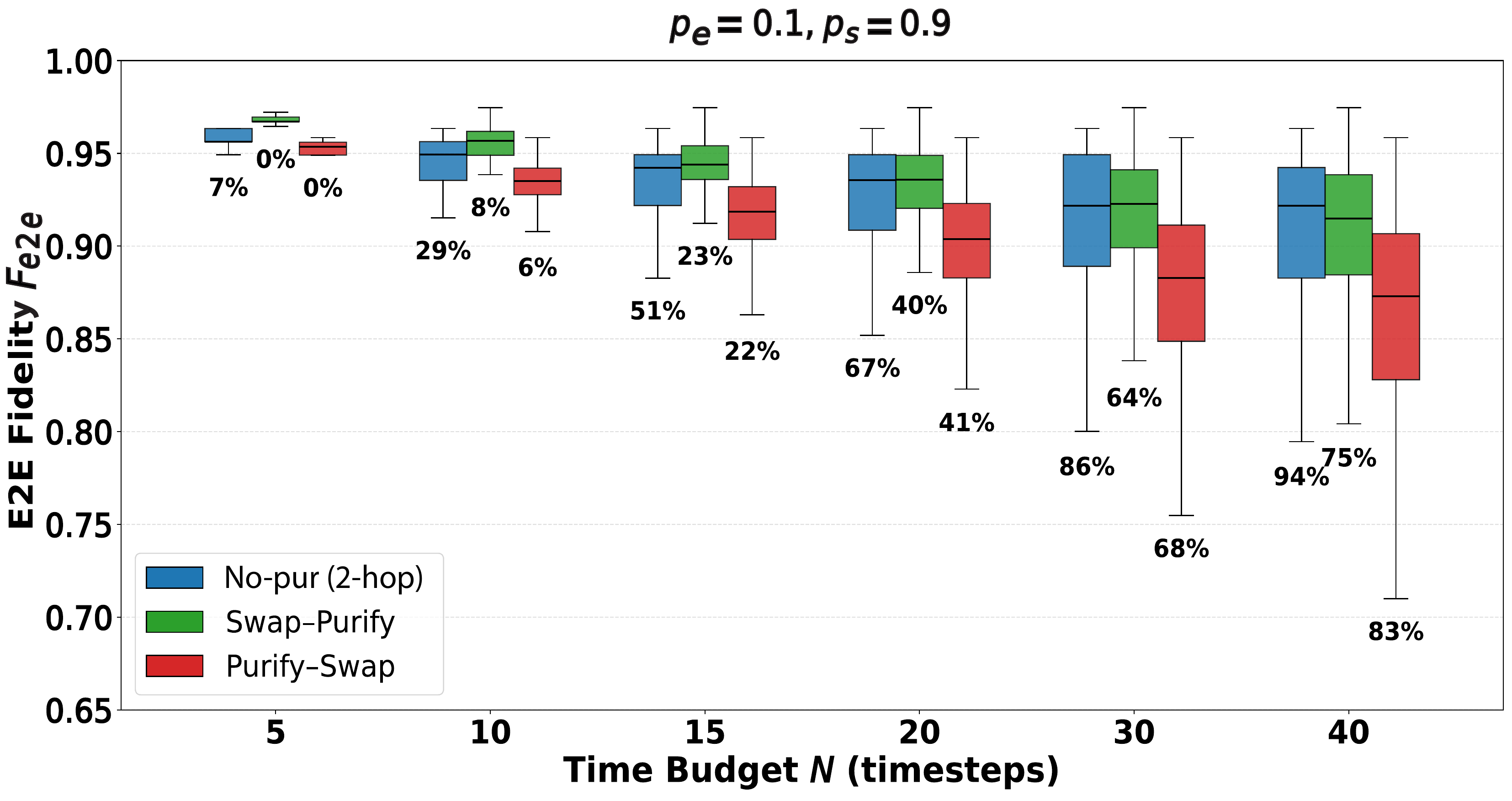}
        \caption{End-to-end fidelity distributions for successfully delivered entanglement across time budgets $N$, with delivery rate $\eta$ annotated below each boxplot.}
        \label{fig:obj2_utility}
    \end{minipage}
\end{figure*}

\begin{figure*}[t]
    \centering
    \includegraphics[width=0.8\linewidth]{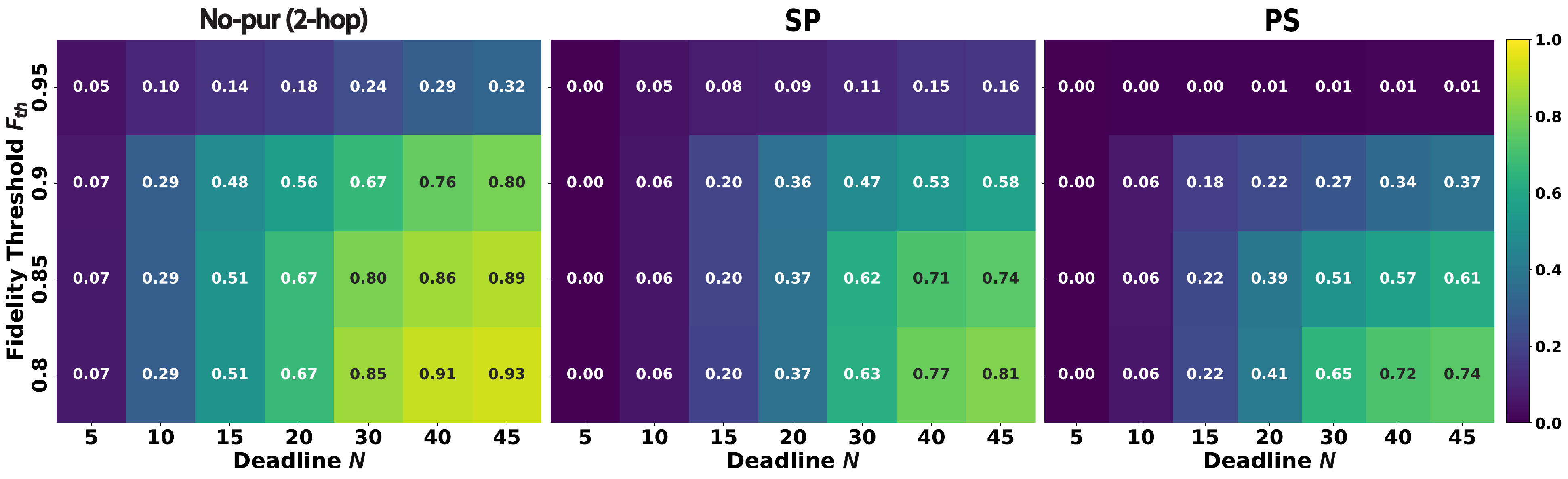}
    \caption{Entanglement delivery rates $\eta \in [0, 1]$ as a function of fidelity threshold $F_{th}$ (y-axis) and time budget $N$ (x-axis) for No-Pur (left), SP (centre), and PS (right). }
    \label{fig:obj3_heatmaps}
\end{figure*}

\subsection{Impact of Memory Model on Purification Gain}
\label{sec:results_memory}

The analysis in Section~\ref{sec:asymmetry} predicts that $\Delta F$ will exceed $\delta(F)$ in the majority of trials under any time-dependent decoherence model, rendering purification counterproductive. We verify this empirically using $\Delta F_{\text{gain}}$ (Eq.~\ref{eq:gain_def}) across the three memory models defined in Section~\ref{subsec:system-definitions}.

Figure~\ref{fig:violin_memory} shows the distribution of $\Delta F_{\text{gain}}$ under CMM, LMM, and EMM. Under CMM, $\Delta F \approx 0$ and purification is unconditionally beneficial across all simulation runs. However, once temporal decoherence is introduced, purification yields a positive gain in only $16.3\%$ of runs under LMM and $14.3\%$ under EMM, confirming that $\Delta F$ exceeds $\delta(F)$ in the majority of trials. The bulk of the distribution of $\delta(F)$ lies below zero, i.e., purification actively degrades the best available input rather than improving it.

Reducing $F_0$ to $0.9$ increases the fraction of beneficial runs to $38\%$ under the EMM and $45\%$ under the LMM; however, the mean gain remains negative in both cases. This confirms that even when a higher fraction of trials fall within the tolerance window, naive purification without awareness of $\delta(F)$ remains the wrong default as the majority of purification attempts still degrade fidelity.

The LMM is included as an optimistic projection of near-term hardware, in which fidelity decays more slowly than under the EMM. That the LMM fails to significantly outperform the EMM confirms that the failure of purification is a structural consequence of asynchronous heralding, not an artifact of current hardware limitations.

\subsection{Policy Comparison and Network Performance}
\label{subsec:policy-comparison}
For all subsequent analyses, we adopt the EMM as it provides the most physically accurate characterization of near-term quantum hardware. While we employ a depolarizing channel model, the results remain qualitatively consistent for other exponential decoherence processes such as combined amplitude damping and dephasing. We assume sufficient memory resources at each node and no resource contention, isolating the effect of fidelity decay due to asynchronous entanglement generation.

We compare three purification policies, No-Pur, SP, and PS, across three objectives that together span the range of application-layer requirements in quantum networks. Given that purification yields a positive gain in only $14.3\%$ of trials under the EMM, No-Pur is a natural candidate, avoiding the overhead and fidelity degradation that naive purification incurs in the majority of cases.

\subsubsection{\textbf{Fidelity-Constrained Delivery}} 

The first objective considers applications that require end-to-end entanglement with fidelity $\geq F_{\text{th}}$, with no constraint on delivery time. In a realistic network, all policies will eventually satisfy the threshold given sufficient time, so delivery rate $\eta = 1$ for all policies under this objective. The distinguishing metric is therefore the time required to do so, and the controller or any node making the purification decision should select the policy that satisfies the threshold in the shortest time.

Figure~\ref{fig:obj1_latency} shows the delivery time distributions across varying $F_{\text{th}}$. No-Pur achieves the lowest median delivery time and narrowest spread across all threshold values. SP requires two complete end-to-end entangled pairs before purification can be attempted, while PS requires two entangled pairs on each elementary link before any swap can proceed. As established in Section~\ref{sec:asymmetry}, the waiting time between generation events drives $\Delta F$ beyond $\delta(F)$ in the majority of trials under the EMM, forcing re-initiation. PS incurs the greatest latency, as this overhead compounds independently at every link before any swap can take place. No-Pur is the optimal purification policy under this objective.

\subsubsection{\textbf{Time-Constrained Delivery}}
The second objective considers applications that impose a strict time budget of $N$ timesteps for end-to-end delivery. Both fidelity and delivery rate $\eta$ are relevant here, but since the application's primary requirement is that entanglement is delivered within $N$, the rate at which this is achieved takes precedence over the fidelity of successful deliveries.

Figure~\ref{fig:obj2_utility} shows the fidelity distributions of successfully delivered pairs across varying $N$, with $\eta$ annotated below each boxplot. For small $N$, SP produces higher fidelity among successful runs than No-Pur. These are the rare trials in which both pairs are heralded quickly enough that $\Delta F < \delta(F)$ is satisfied, allowing purification to yield a positive gain. In practice, these occurrences are exceedingly rare, as reflected by the negligibly small $\eta$ for SP at low $N$. As $N$ increases and $\eta$ grows, the trials contributing to the SP distribution are those in which pairs spent longer in memory, and the resulting asymmetry drags the mean fidelity down as negative-gain cases come to dominate. Achieving high fidelity through SP therefore requires an application willing to accept a very low delivery rate, which renders SP impractical for applications with any meaningful throughput requirement. No-Pur delivers lower peak fidelity but satisfies the time budget in substantially more trials across all values of $N$, which is the relevant criterion under this objective.

This result illustrates the role of $F_{\lim}$ (Eq.~\ref{eq:flim}), the optimistic upper bound on end-to-end fidelity achievable without purification, beyond which No-Pur cannot deliver regardless of timing. The full derivation of $F_{\lim}$ accounting for memory decay and stochastic generation is out of the scope of this work. As an optimistic upper bound, we apply Eq.~\ref{eq:swap} recursively across $n$ hops under the assumption of no memory decay, giving
\begin{equation}
\label{eq:flim}
F_{\lim} = \frac{1}{4} + \frac{3}{4}\prod_{i=1}^{n}
\left(\frac{4F_i - 1}{3}\right).
\end{equation}

Since memory decay during swap-ASAP generation can only reduce the achievable fidelity below  $F_{\lim}$, the condition $F_{\text{th}} > F_{\lim}$ is a sufficient condition for purification to be necessary. If $F_{\text{th}} < F_{\lim}$, the results above confirm that No-Pur is the superior choice.

\subsubsection{\textbf{Joint-Constrained Delivery}}

The third objective evaluates performance when the application specifies both $F_{\text{th}}$ and $N$ simultaneously, making it the most demanding of the three objectives considered. Policies are compared on delivery rate $\eta(F_{\text{th}}, N)$, the fraction of runs delivering entanglement with fidelity $\geq F_{\text{th}}$ within $N$ timesteps.

Figure~\ref{fig:obj3_heatmaps} shows $\eta(F_{\text{th}}, N)$ for each policy across the joint parameter space. No-Pur sustains the largest high-rate region, maintaining elevated $\eta$ across a broader range of $(F_{\text{th}}, N)$ combinations than either SP or PS. Purification-based policies achieve comparable rates only in the high-$N$, low-$F_{\text{th}}$ regime where both constraints are relaxed. That No-Pur remains the optimal purification policy even under this joint constraint confirms that the performance gap between purification-based policies and No-Pur is not specific to any single network objective, but is a consistent consequence of the asymmetry-induced fidelity degradation.

\begin{figure*}[t]
    \centering
    \begin{minipage}[t]{0.48\linewidth}
        \centering
        \includegraphics[width=\linewidth]{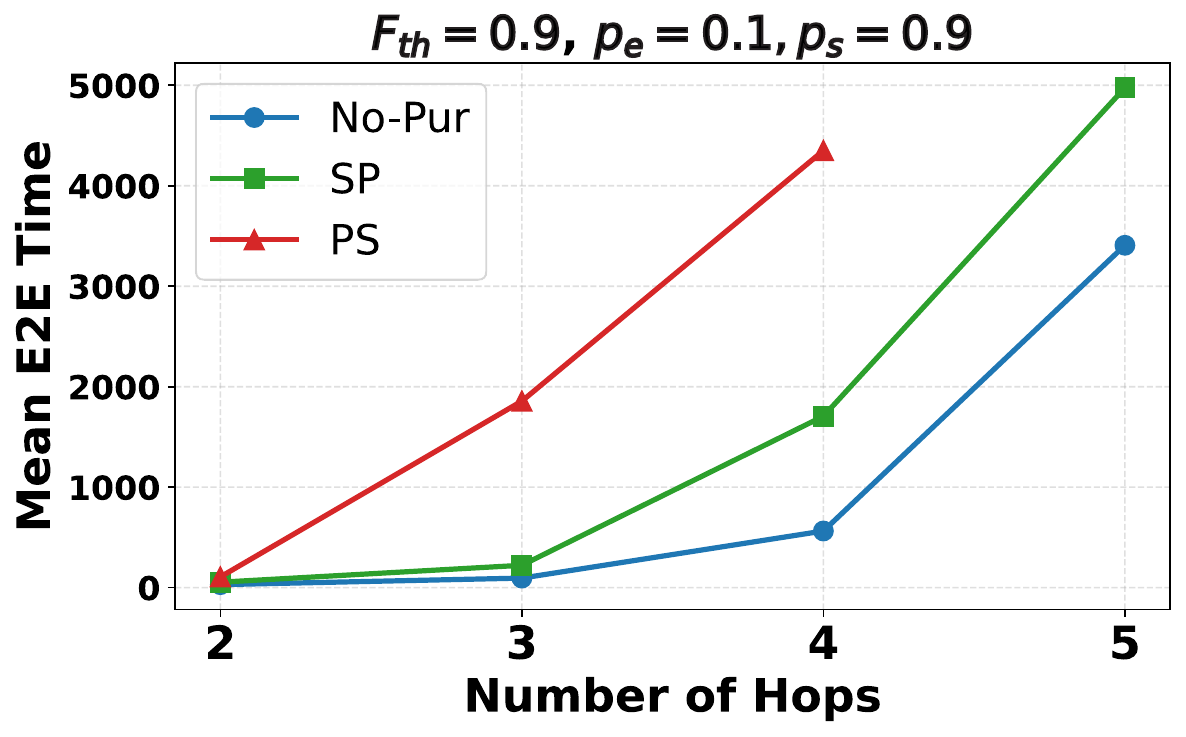}
        \caption{Mean end-to-end delivery time as a function of hop count for No-Pur, SP, and PS at $F_{\text{th}} = 0.9$. No-Pur grows moderately with hop count while SP and PS grow substantially more steeply. The PS curve does not extend to 5 hops, as delivery within the simulation cutoff of $10{,}000$ timesteps was not achieved at that scale.}
        \label{fig:obj1_scalability}
    \end{minipage}
    \hfill
    \begin{minipage}[t]{0.48\linewidth}
        \centering
        \includegraphics[width=\linewidth]{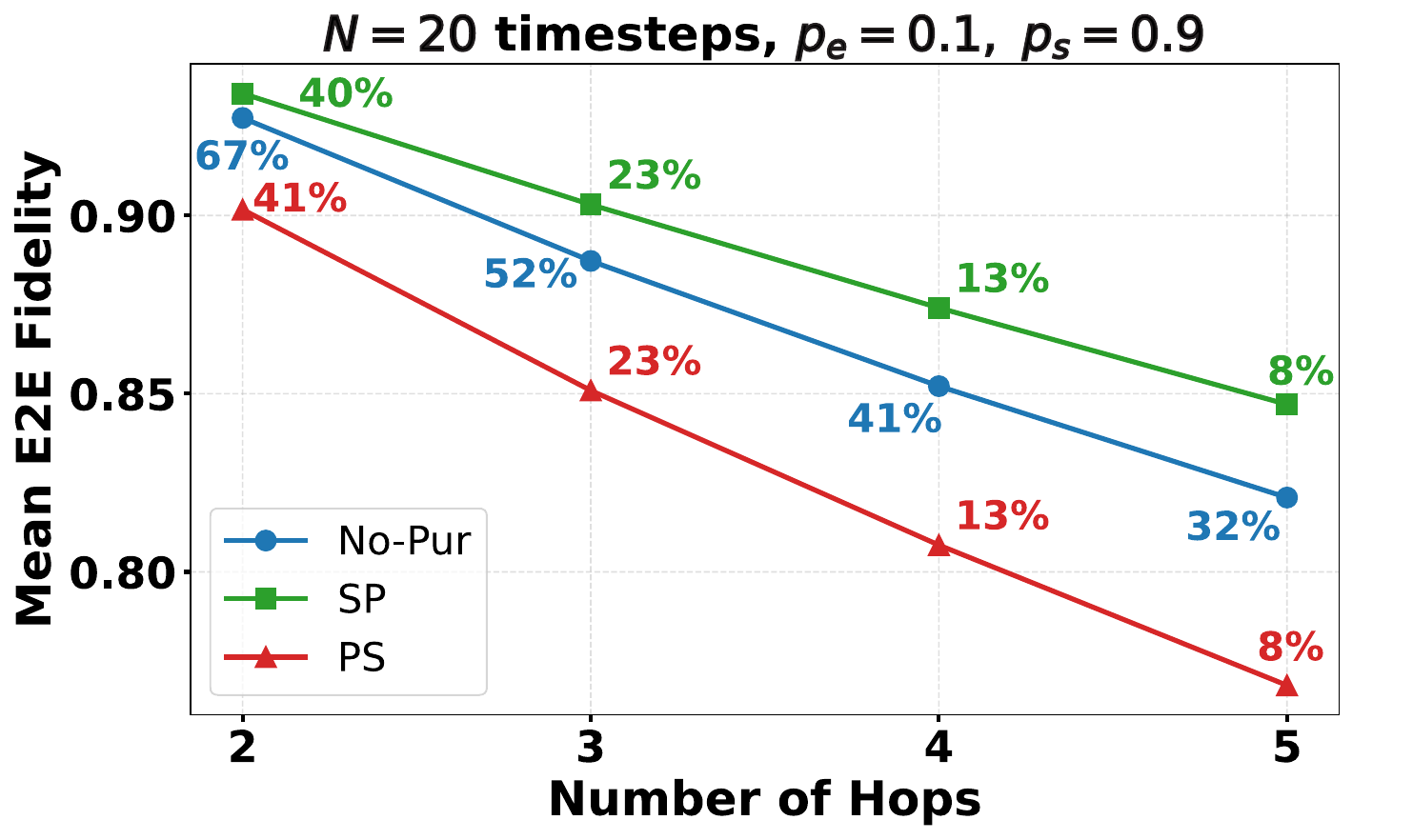}
        \caption{Mean end-to-end fidelity of successfully delivered pairs as a function of hop count at $N = 20$, with delivery rate $\eta$ annotated for each policy. SP achieves the highest mean fidelity but its delivery rate declines steeply with hop count. No-Pur maintains competitive fidelity with a substantially higher and more stable delivery rate across all hop counts tested.}
        \label{fig:obj2_scalability}
    \end{minipage}
\end{figure*}

\begin{figure*}[t]
    \centering
    \includegraphics[width=\linewidth]{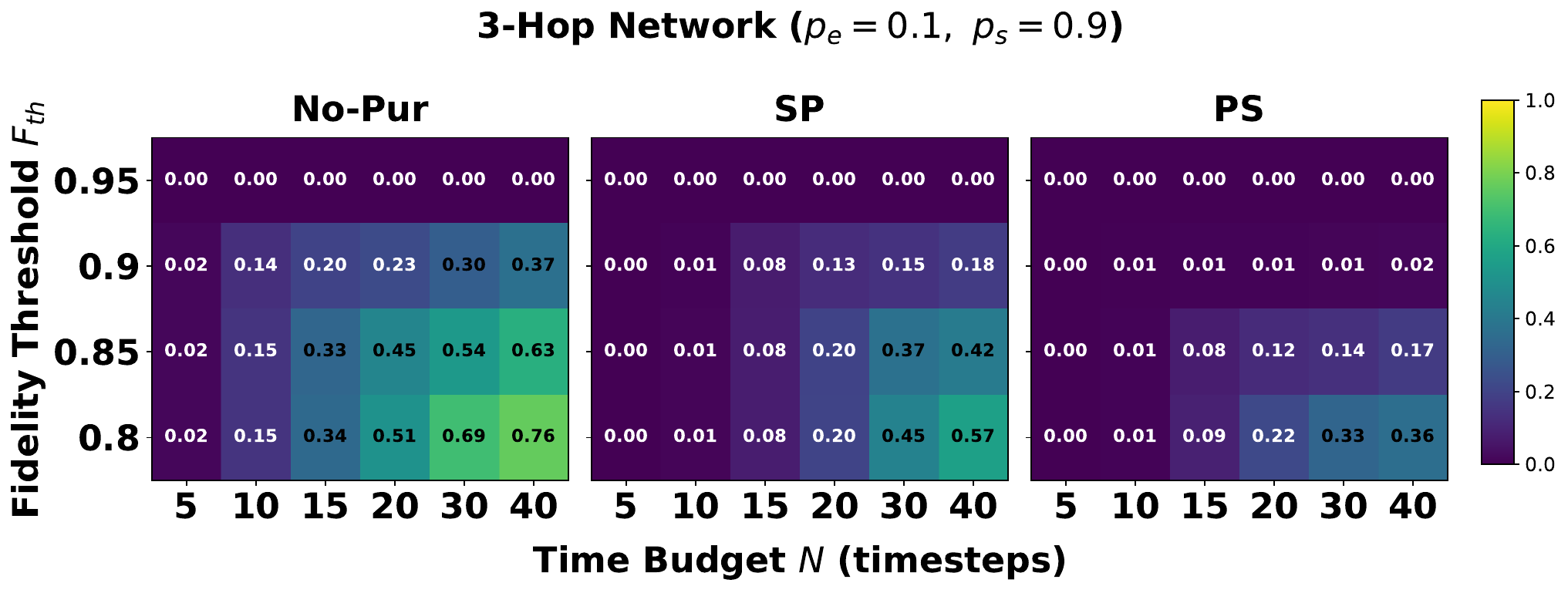}
    \caption{Entanglement delivery rate $\eta \in [0, 1]$ as a function of $F_{\text{th}}$ (y-axis) and $N$ (x-axis) for No-Pur (left), SP (centre), and PS (right), for $n = 3$ hops. No-Pur maintains the widest feasible operating region across the joint parameter space. The relative ordering of policies is preserved compared to the 2-hop case (Fig.~\ref{fig:obj3_heatmaps}).}
    \label{fig:obj3_scalability}
\end{figure*}

\subsection{Scalability with Number of Hops}
\label{subsec:scalability}

We examine how the performance changes with network size by extending the repeater chain from 2 to 5 hops. The conclusions hold across the full range of $F_{\text{th}}$ and $N$ values tested.

\subsubsection{\textbf{Fidelity-Constrained Delivery}}

Figure~\ref{fig:obj1_scalability} shows the mean end-to-end delivery time as a function of hop count for $F_{\text{th}} = 0.9$. Under No-Pur, delivery time grows moderately with hop count. For SP and especially PS, it grows substantially more steeply, as the purification overhead compounds across hops. \textbf{No-Pur's advantage over SP and PS therefore widens as the hop count increases.}
For 5 hops and $F_{\text{th}} = 0.9$, PS fails to deliver within the simulation cutoff of $10{,}000$ timesteps at this scale. This absence is itself informative: the cumulative purification overhead at every link in a five-hop chain renders PS infeasible at this combination of network size and fidelity requirement.\\

\subsubsection{\textbf{Time-Constrained Delivery}}

Figure~\ref{fig:obj2_scalability} shows the mean end-to-end fidelity of successfully delivered pairs as a function of hop count for $N = 20$, with $\eta$ annotated for each policy. SP achieves the highest mean fidelity across all hop counts, but its delivery rate $\eta$ falls substantially below that of No-Pur and declines more steeply as hop count increases. As discussed in Section~\ref{subsec:policy-comparison}, SP's higher fidelity reflects the few rare trials where decoherence has not yet driven $\Delta F$ beyond $\delta(F)$. As the number of hops increases, decoherence accumulates, shrinking these rare trials further. No-Pur maintains a competitive mean fidelity with a substantially higher and more stable delivery rate across all hop counts. PS produces the lowest fidelity and lowest delivery rate at every hop count. Link-level purification  consumes a disproportionate share of the time budget. No-Pur is therefore the best purification policy under this objective and its advantage strengthens as the network grows.\\

\subsubsection{\textbf{Joint-Constrained Delivery}}

Figure~\ref{fig:obj3_scalability} shows $\eta(F_{\text{th}}, N)$ for a representative topology of $n = 3$ hops. No-Pur maintains the widest feasible operating region across the joint parameter space, and SP and PS rarely reach comparable delivery rates. As hop count increases, the feasible operating region contracts for all three policies, but the contraction is more severe for SP and PS. The relative ordering of policies is preserved across all hop counts tested, and the conclusions established for the 2-hop topology in Section~\ref{subsec:policy-comparison} generalize (No-Pur is best) as the hop count increases.

\section{The \texorpdfstring{$\delta$}{delta}-Informed Purification Policy}
\label{sec:deltapurify}

The results of Section~\ref{sec:results} reveal two distinct operating regimes. When $F_{\text{th}} \leq F_{\lim}$, No-Pur is superior and when $F_{\text{th}} > F_{\lim}$, purification is structurally necessary. A rigid policy cannot handle both regimes effectively. Naive purification incurs unnecessary overhead in the first regime, and No-Pur is infeasible in the second. The $\delta(F)$ criterion derived in Section~\ref{sec:asymmetry} provides the analytical foundation for a policy that handles both regimes by evaluating the purification decision dynamically based on the actual state of the network at the moment the decision must be made. We present the $\delta$-informed purification policy, DeltaPurify, which uses $F_{\lim}$ and $\delta(F)$ to make proactive, state-dependent decisions at each stage of entanglement distribution.

\subsection{Policy Description}

Upon receiving a request with threshold $F_{\text{th}}$ on a selected
$n$-hop path, DeltaPurify first compares $F_{\text{th}}$ against $F_{\lim}$
(Eq.~\ref{eq:flim}); this check uses only the path's link fidelities and can
be done before any resources are committed. If $F_{\text{th}} \leq
F_{\lim}$, purification is skipped and generation proceeds swap-ASAP until a
pair with fidelity $\geq F_{\text{th}}$ is obtained. Otherwise, once the
first end-to-end pair (fidelity $F_1$) is generated, a feasibility check
evaluates the best purified fidelity achievable with $F_1$ as one input ---
either a symmetric second pair ($F_2 = F_1$) or one at the upper limit of
the $\delta$ criterion, $F_2 = \min(F_1+\delta(F_1),1)$
(Eq.~\ref{eq:delta_unified}) --- via
\begin{align}
\hat{F}_{\text{pur}}^{A} &= F_{\text{pur}}(F_1, F_1), \\
\hat{F}_{\text{pur}}^{B} &= F_{\text{pur}}(\min(F_1 + \delta(F_1), 1), F_1), \\
\hat{F}_{\text{pur}} &= \max\left(\hat{F}_{\text{pur}}^{A}, \hat{F}_{\text{pur}}^{B}\right).
\end{align}
If $\hat{F}_{\text{pur}} < F_{\text{th}}$, $F_1$ is discarded and generation
restarts, avoiding a doomed attempt. Otherwise, generation of the second
pair proceeds; once both fidelities are known, the observed asymmetry
$\Delta F = |F_1' - F_2|$ is checked against $\delta(F)$
(Eq.~\ref{eq:delta_unified}), and purification is attempted only if $\Delta
F < \delta(F)$, with both pairs discarded and generation restarted
otherwise. Discarded pairs could in principle be reassigned to other
concurrent requests; we assume isolated single-demand operation and leave
multi-demand reuse to future work. Algorithm~\ref{alg:deltapurify}
formalizes this procedure.

\begin{algorithm}[ht]
\caption{$\delta$-Informed Purification Policy 
(DeltaPurify)}
\label{alg:deltapurify}
\textbf{Input:} Hops $n$, fidelity threshold 
$F_{\text{th}}$, system parameters $\mathcal{P}$\\
\textbf{Output:} Completion time $T$, final fidelity $F$
\begin{algorithmic}[1]
\If{$F_{\text{th}} \leq F_{\lim}$}
    \Repeat
        \State Run \textsc{GenE2E} $\Rightarrow$ 
        obtain $(t_1, F_1)$
        \State $T \mathrel{+}= t_1$
        \If{$F_1 \geq F_{\text{th}}$}
            \State \Return $(T, F_1)$
        \EndIf
    \Until{success}
\Else
    \Repeat
        \Statex \textbf{// Phase 1: Generate first 
        end-to-end pair}
        \State Run \textsc{GenE2E} $\Rightarrow$ 
        obtain $(t_1, F_1)$
        \State $T \mathrel{+}= t_1$
        \Statex \textbf{// Feasibility check}
        \State $\hat{F}_{\text{pur}}^A \leftarrow F_{\text{pur}}(F_1, F_1)$
        \State $\hat{F}_{\text{pur}}^B \leftarrow F_{\text{pur}}\!\left(\min(F_1 + \delta(F_1), 1),\; F_1\right)$
        \State $\hat{F}_{\text{pur}} \leftarrow \max\!\left(\hat{F}_{\text{pur}}^A,\; \hat{F}_{\text{pur}}^B\right)$
        \If{$\hat{F}_{\text{pur}} < F_{\text{th}}$}
            \State Discard $F_1$
            \State \textbf{continue}
        \EndIf
        \Statex \textbf{// Phase 2: Generate second 
        end-to-end pair}
        \State Run \textsc{GenE2E} $\Rightarrow$ 
        obtain $(\Delta t, F_2)$
        \State Apply decoherence to $F_1$ for 
        $\Delta t$ timesteps $\Rightarrow F_1'$
        \State $T \mathrel{+}= \Delta t$
        \Statex \textbf{// $\delta$ check}
        \State $\Delta F \leftarrow |F_1' - F_2|$
        \If{$\Delta F \geq 
        \delta(F)$ \textbf{// (Eq.~\ref{eq:delta_unified}})}
            \State Discard both
            \State \textbf{continue}
        \EndIf
        \Statex \textbf{// Phase 3: Attempt 
        purification}
        \State $(success, F_{\text{pur}}) \leftarrow 
        \textsc{Purify}(F_1', F_2)$
        \State $T \mathrel{+}= 1$
        \If{$success$ \textbf{ and } $F_{\text{pur}} 
        \geq F_{\text{th}}$}
            \State \Return $(T, F_{\text{pur}})$
        \EndIf
        \State Discard both
    \Until{success}
\EndIf
\end{algorithmic}
\end{algorithm}

\subsection{Performance under Fidelity-Constrained Delivery}

DeltaPurify is designed to make proactive decisions early in the generation process, avoiding resource expenditure on purification attempts that are guaranteed to fail or degrade fidelity. These early decisions are expected to reduce the time required to serve a request. We therefore evaluate DeltaPurify against No-Pur and SP on the time to serve, defined as the time required to deliver end-to-end entanglement with fidelity $\geq F_{\text{th}}$, across varying threshold values. SP is included as the reference purification-based policy given its superior performance over PS in the results of Section~\ref{sec:results}.

\begin{figure}[t]
    \centering
    \includegraphics[width=0.8\linewidth]{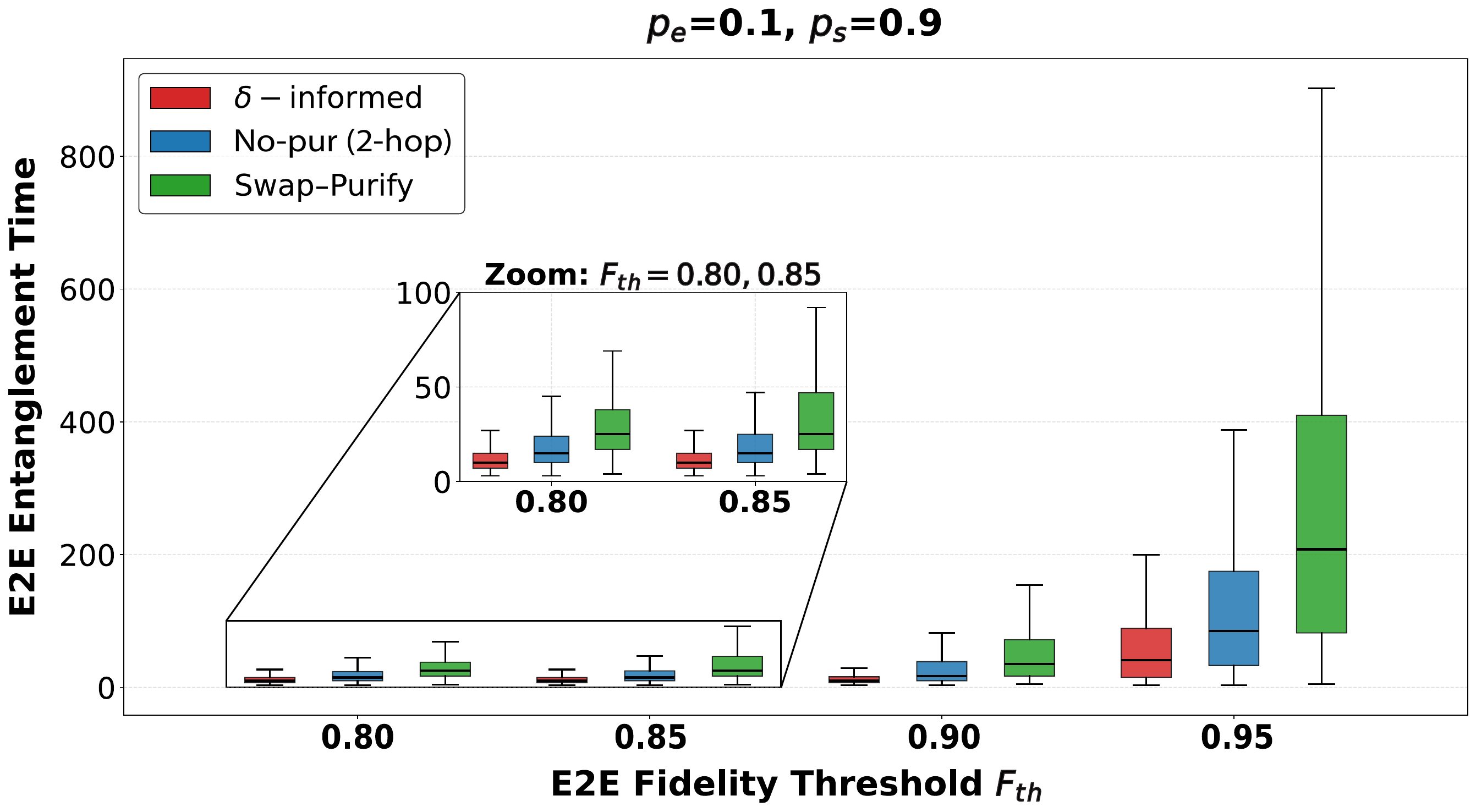}
    \caption{Distribution of time to serve for No-Pur, 
    SP, and DeltaPurify across fidelity thresholds 
    $F_{\text{th}}$ for a 2-hop topology. DeltaPurify 
    achieves the lowest median and narrowest spread 
    across all threshold values, demonstrating that 
    $\delta$-informed purification decisions reduce 
    delivery latency relative to both naive 
    purification and no-purification.}
    \label{fig:deltapurify_obj1}
\end{figure}

Figure~\ref{fig:deltapurify_obj1} shows the distribution of time to serve across varying $F_{\text{th}}$ for a 2-hop topology. DeltaPurify achieves substantially lower time to serve than both No-Pur and SP across all threshold values. The reduction relative to No-Pur arises because DeltaPurify exploits purification when the $\delta(F)$ condition is satisfied, reaching higher fidelity thresholds more quickly than No-Pur alone. The reduction relative to SP arises because DeltaPurify avoids two sources of wasted overhead. When the feasibility check or $\delta$ check determines that purification will not help, generation is restarted early rather than running each attempt to completion. When $F_{\text{th}} \leq F_{\lim}$, purification is bypassed entirely, and delivery proceeds without the overhead of generating additional pairs.

\begin{figure}[t]
    \centering
    \includegraphics[width=0.8\linewidth]{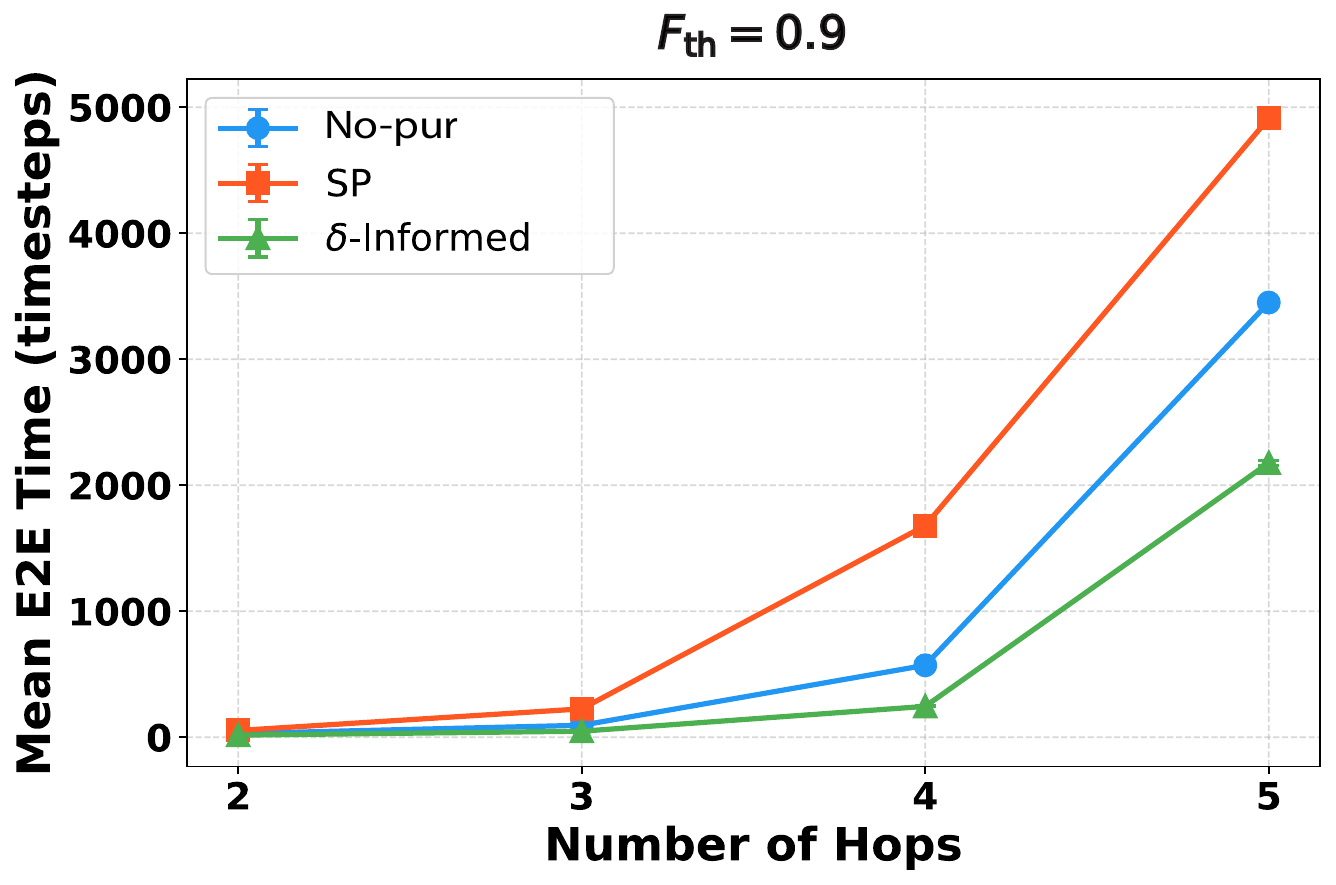}
    \caption{Mean time to serve as a function of hop 
    count for No-Pur, SP, and DeltaPurify at 
    $F_{\text{th}} = 0.9$. DeltaPurify achieves the 
    lowest mean time to serve across all hop counts 
    tested, and its advantage over both No-Pur and 
    SP is preserved across all network sizes 
    considered.}
    \label{fig:deltapurify_scalability}
\end{figure}

Figure~\ref{fig:deltapurify_scalability} shows the mean time to serve as a function of hop count at $F_{\text{th}} = 0.9$. DeltaPurify achieves the lowest mean time to serve across all hop counts tested, and its advantage over both No-Pur and SP is preserved across all network sizes considered. As hop count grows, the benefit of early loss mitigation becomes more pronounced, since the cost of a failed purification attempt that could have been avoided compounds across multiple links.


The $F_{\lim}$ check requires knowledge of the selected path and its link fidelities, available once routing has been determined. All subsequent decisions use only local state information: the fidelities of buffered pairs and their elapsed storage time. If the path changes, $F_{\lim}$ is recomputed from the new link fidelities and the decision logic applies without modification.

\section{Discussion and Future Direction}
DeltaPurify operates over a single round of two-copy purification and
extends naturally to multiple rounds by applying $\delta(F)$ iteratively; as
path length grows and swapping-only fidelity drops, the number of resource
pairs needed depends not just on success probability but on the fidelity
gap between pairs --- a dependence hidden under the symmetric-input
assumption. As shown in Section~\ref{sec:asymmetry}, this asymmetry analysis
extends unchanged to DEJMPS and other two-copy CNOT-recurrence protocols on
Werner inputs; extending it to $r$-to-1 distillation protocols with $r>2$
\cite{jansen2022enumerating}, where coordinating more stochastic generation
processes worsens the asymmetry problem, and to multi-round nested
purification, is left to future work.

A tighter $F_{\lim}$ accounting for memory decoherence during swap-ASAP
generation is available from Goodenough et
al.~\cite{goodenough2025noise} and, combined with Brand et
al.~\cite{brand2020efficient} or the sequential-swapping expressions of
Guedes de Andrade et al.~\cite{de2024analysis}, could enable a
proactive DeltaPurify that allocates resources at request arrival time; a
first-completion policy~\cite{fayyaz2025selecting} offers a complementary
strategy when $F_{\text{th}} \leq F_{\lim}$. Our evaluation is restricted to
linear repeater chains with a single demand and no resource contention;
extending it to general topologies and multi-demand traffic, and evaluating
resource-cost metrics such as memory occupation time, per-pair resource
efficiency, and fairness among requests, are important directions for
future work.

\section{Conclusion}
We analyzed the conditions under which entanglement purification is beneficial in a quantum repeater chain under realistic memory decoherence, showing that input fidelity asymmetry renders naive purification counterproductive in the majority of trials. We derived a closed-form asymmetry tolerance $\delta_{\max} \approx 0.076$, and showed that when the application fidelity threshold is achievable through swapping alone, no-purification is the superior policy with its advantage widening with network size. Building on these results, DeltaPurify conditions purification decisions on local fidelity information and reduces time-to-serve relative to both naive purification and no-purification across all thresholds and network sizes tested.

\section*{Acknowledgment}
KPS thanks the U.S. Department of Energy, Office of Science, Advanced Scientific Computing Research (ASCR) program, for support under Award Number DE-SC0026264. KPS and PK thank Pittsburgh Quantum Institute Community Collaboration Awards.

\end{document}